\documentclass[10pt,letterpaper]{article}
\usepackage[top=0.85in,left=2.75in,footskip=0.75in]{geometry}

\usepackage{amsmath,amssymb}

\usepackage{changepage}

\usepackage[utf8x]{inputenc}

\usepackage{textcomp,marvosym}

\usepackage{cite}

\usepackage{nameref,hyperref}

\usepackage[right]{lineno}

\usepackage{microtype}
\DisableLigatures[f]{encoding = *, family = * }

\usepackage[table]{xcolor}

\usepackage{array}

\newcolumntype{+}{!{\vrule width 2pt}}

\newlength\savedwidth

\newcommand\thickhline{\noalign{\global\savedwidth\arrayrulewidth\global\arrayrulewidth 2pt}%
\hline
\noalign{\global\arrayrulewidth\savedwidth}}

\raggedright
\usepackage[aboveskip=1pt,labelfont=bf,labelsep=period,justification=raggedright,singlelinecheck=off]{caption}

\makeatletter
\renewcommand{\@biblabel}[1]{\quad#1.}
\makeatother

\usepackage{lastpage,fancyhdr,graphicx}
\usepackage{epstopdf}
\fancyheadoffset[L]{2.25in}
\fancyfootoffset[L]{2.25in}
\begin{document}
\vspace*{0.2in}

\begin{flushleft}
{\Large
\textbf\newline{Ceibaco: REST API and Single Page Application for the generation and evaluation of bijective S-boxes} 
}
\newline
\\
Ismel Martínez-Díaz\textsuperscript{1*}
\\
\bigskip
\textbf{1} Ceibaco Lab, Havana, Cuba

\bigskip

*diomedes.martnezdaz2@gmail.com

\end{flushleft}
\section*{Abstract}
In this paper we present the first REST API for the generation and evaluation of bijective S-boxes. We also present the first Single Page Application tool for researchers and students that allows the use of a graphical interface. We give a small dataset of classical S-boxes to test the properties evaluations. We show how to define experiments and we include two local search experiments into the proposed tool.



\section*{Introduction}

 The S-box is a primary component of block ciphers and it is essential for keeping the confusion in encryption/decryption process \cite{van2014encyclopedia}. The evaluation of an S-box corresponds to the values of several properties that reflect the S-box's resistance against attacks. Some of the most important properties are Balance \cite{van2014encyclopedia}, Non-Linearity (NL) \cite{nyberg1992construction, carlet2007nonlinearities}, Differential Uniformity (DU) \cite{nyberg1993differentially, carlet2010vectorial}, Confusion Coefficient Variance (CCV) \cite{picek2014confused} and Transparency Order  \cite{chakraborty2017redefining, li2020notion}.
 
 The S-boxes can be generated. Main methods are random generation - including chaos-based - \cite{lambic2013comparison, alkhayyat2022novel, abduljabbar2022provably}, algebraic construction \cite{nyberg1993differentially, bergman2022constructing, jeon2022differential, shah2022cryptographically} and evolutionary computation \cite{picek2015applications, freyre2020evolving, behera2021evolving}. There also exist hybrid methodologies  \cite{jimenez2017generation, carlet2022evolving}.
 
 In the actual days there are some tools for the generation and evaluations of S-boxes. The S-box Evaluation Tool (SET) and the  Generation and Evaluation Tool (GET) were presented by Picek~\cite{picek2015applications}; both were developed using the \textbf{C} programming language. In the case of the GET tool, it includes generation using evolutionary computation experiments. Recently, Banmagrupta \textit{et al.} ~\cite{behera2021bsat} create a new \textbf{C++} program: Boolean function and S-box Analysis Tool (BSAT), improving the speed of properties evaluation. A graphical tool is presented in \cite{ibrahim2021random} for the chaotic-based random generation of S-boxes, this tool includes the evaluation of diffusion properties like Strict Avalanche Criterion and Bit Independence Criterion \cite{van2014encyclopedia}; it was conceived under the \textbf{C\#} programming language.

Most of actual tools has been developed in \textbf{C/C++} languages, because they provide good performance and portability across operating systems. SET, GET and BSAT are Command Line Interface applications that can be executed into a terminal emulator. Banmagrupta \textit{et al.} in ~\cite{behera2021bsat} pointed out the convenience of having a graphical user interface tool. The application presented in \cite{ibrahim2021random}, despite being graphical, used the weak definition of Non-Linearity, where take relevance the average of the Non-Linearity of the coordinate boolean functions that belongs to the S-box. 

Today, applications are looking for accessibility across all platforms and devices. The web applications or web-apps are great for this goal, and in a particular case a Single Page Application (SPA) \cite{fink2014pro, spa2019article}. Also  REST API \cite{subramanian2019hands} provide to all developers the possibility of use different programming languages and platforms to create new applications - experiments in the case of researchers -.

In this paper we present the first REST API for generate an evaluate S-boxes. We also present a new web tool for researchers and students. The paper is organized as follow: in the next section are shown the main materials and methods used to build and conform the tools; in the Results section we describe the software architecture of the web tool and also the experiments and tests. Finally, we continue with the Discussion and Conclusion sections, and give new research directions.  

\section*{Materials and methods}

In this section we present the essential notions of S-boxes properties. We show the actual state of the REST API and Single Page Applications technologies.

\subsection*{S-box properties}

An S-box is a mapping from the vector space $\lbrace 0,1 \rbrace^{n}$ into the vector space $\lbrace 0,1 \rbrace^{m}$, $n,m$ positive integer numbers. It also can be seen as a vector of its coordinate boolean functions $F=(f_1,...,f_m)$, where a boolean function $f: \lbrace 0,1 \rbrace^{n} \rightarrow \lbrace 0,1 \rbrace $ makes the correspondence between \textit{n} input bytes with one output bit. In the particular case of actual block ciphers, it is common that $m=n$. In this work we denote an S-box as a bijective vector boolean function $F: \lbrace 0,1 \rbrace^{n} \rightarrow \lbrace 0,1 \rbrace^{m}, m = n$.\\

There are some main S-box properties. For our application, we take into account the confusion coefficient variance, the transparency order, the non-linearity, the delta uniformity and the balance.

\subsubsection*{Side-Channel related properties}
In \cite{picek2014confused} is proposed the Confusion Coefficient Variance (CCV) using the Hamming Weight power model to simulate the leakages. It's formula, for all sub-keys $k_i, k_j, k_i \neq k_j$ and all input text $in$, is Eq~(\ref{eq:vcc}):
\begin{equation}
\label{eq:vcc}
CCV(F) = Var(E[(HW(F(in \oplus k_i)) - HW(F(in \oplus k_j)))^2])
\end{equation}\\

In \cite{chakraborty2017redefining},  Modified Transparency Order (MTO) takes into account the cross-correlation spectrum of the coordinate functions of the S-box $F=(f_1,...,f_m)$, denoted by 

\begin{equation*}
C_{f_i, f_j}(\alpha) = \sum_{ x \in \lbrace 0,1 \rbrace^{n}  }{(-1)^{  f_i(x) \oplus f_j(x \oplus \alpha) }}
\end{equation*}

The property is computed as:
\begin{equation}
\label{eq:mto}
MTO(F) = max_{ \beta \in \lbrace 0,1 \rbrace^{m}  } ( m - \frac{1}{2^{2^{n}} - 2^{n}} \sum_{\alpha \in \lbrace 0,1 \rbrace^{n} - \lbrace 0 \rbrace^{n}} \sum_{j = 1}^{m}   {| \sum_{i = 1}^{m} { (-1)^{\beta_{i} \oplus  \beta_{j} } C_{F_i, F_j}(\alpha) } | } )
\end{equation}\\

In the same fashion of MTO, the Revised Transparency Order is computed as follow:
\begin{equation}
\label{eq:rto}
RTO(F) =  max_{ \beta \in \lbrace 0,1 \rbrace^{m}  } ( m - \frac{1}{2^{2^{n}} - 2^{n}} \sum_{\alpha \in \lbrace 0,1 \rbrace^{n} - \lbrace 0 \rbrace^{n}} | \sum_{j = 1}^{m}   { \sum_{i = 1}^{m} { (-1)^{\beta_{i} \oplus  \beta_{j} } C_{F_i, F_j}(\alpha) }  }| )
\end{equation}\\

\subsubsection*{Algebraic related properties}
The Walsh-Hadamard transform of a boolean function $f$ is defined as
\begin{equation}
WH_f(w) = \sum_{x \in F_2^n}\hat{f}(x)\hat{L_w}(x)
\end{equation}
where $\hat{f}$ represents the polar form of boolean function $f$ and $\hat{L_w}$ is a linear function specified by \textit{w}. The maximum value of Walsh-Hadamard transform for a boolean function $f$ is denoted by
\begin{equation}
WH_{max}(f) = max_{w \in F_2^n} | WH_f(w)|
\end{equation}
in which  represents the absolute value.

The non-linearity (\textbf{NL}) of a boolean function $f$ is defined as 
\begin{equation}
NL_f = \frac{1}{2}(2^n - WH_{max}(f))
\end{equation}	
Then the non-linearity of an S-Box is the lowest value of non-linearity among its component functions (non-zero linear combinations of the $m$ coordinates). 

The differential uniformity of a bijective S-Box $F$ is calculated as follows
\begin{equation}
\delta_F = max_{a, b \in \lbrace 0,1 \rbrace^{n}, a\neq \lbrace 0 \rbrace^{n}} \delta(a, b)
\end{equation}
where $\delta(a, b) = \arrowvert \{x \in \lbrace 0,1 \rbrace^{n} \arrowvert F(x\oplus a)\oplus F(x) = b\}\arrowvert$.\\
\paragraph{Remark 1} The Hamming-Weight (\textbf{HW}) of a vector $\alpha \in F_2^N$ is the number of non-zero positions in the vector. Two S-Boxes $\Phi_1, \Phi_2$ are within the same Hamming-Weight class if $\forall x \in F_2^N$ HW($\Phi_1$(x))=HW($\Phi_2$(x)).

\subsubsection*{Statistical related properties}

All bijective S-boxes are balanced.

\subsection*{REST API}

The REST API is the main type of API used over the HTTP protocol. It provides a way of transfer application state and data between the web client and the web server.

The main methods for state transfer are GET, POST, PUT and DELETE. However, the most used methods are GET and POST, because they can simulate the other two.

The GET method is frequently used to recover resources from the server. The resource type is specified in the url, including search parameters like count, pagination, or particular properties. The response is generally in JavaScript Object Notation (JSON) format.

The POST method is frequently used to create or update one or more resources into the server. The request body is generally in a JSON format, and the response status indicates if the actions was successful or not.

The urls defined for GET and POST are called the endpoints of the application. In our tool we set some endpoints and in the next section we will describe the action of each one.

\subsection*{Single page application}

A single page application (SPA) is an actual form of build web apps. It relies in the use of only one page instead of using several pages as a rich web site app does.

The app is structured as components and services. Most recognized frameworks that can build SPA are React, Angular and Vue. React is a set of decoupled javascript libraries that together define a framework. Angular, in the other hand, is a full typescript-frontend framework that provides all components and services to build an app. Finally, Vue is also a set of libraries like React but tends to present the template language and project structure in the way that Angular does. 

\section*{Results}

In this section we present the main results of our research: the web tool architecture, the small test dataset, the API experiments and the Local Search experiments.

\subsection*{Software architecture}

\begin{figure}[!h]
	\includegraphics[width=\linewidth]{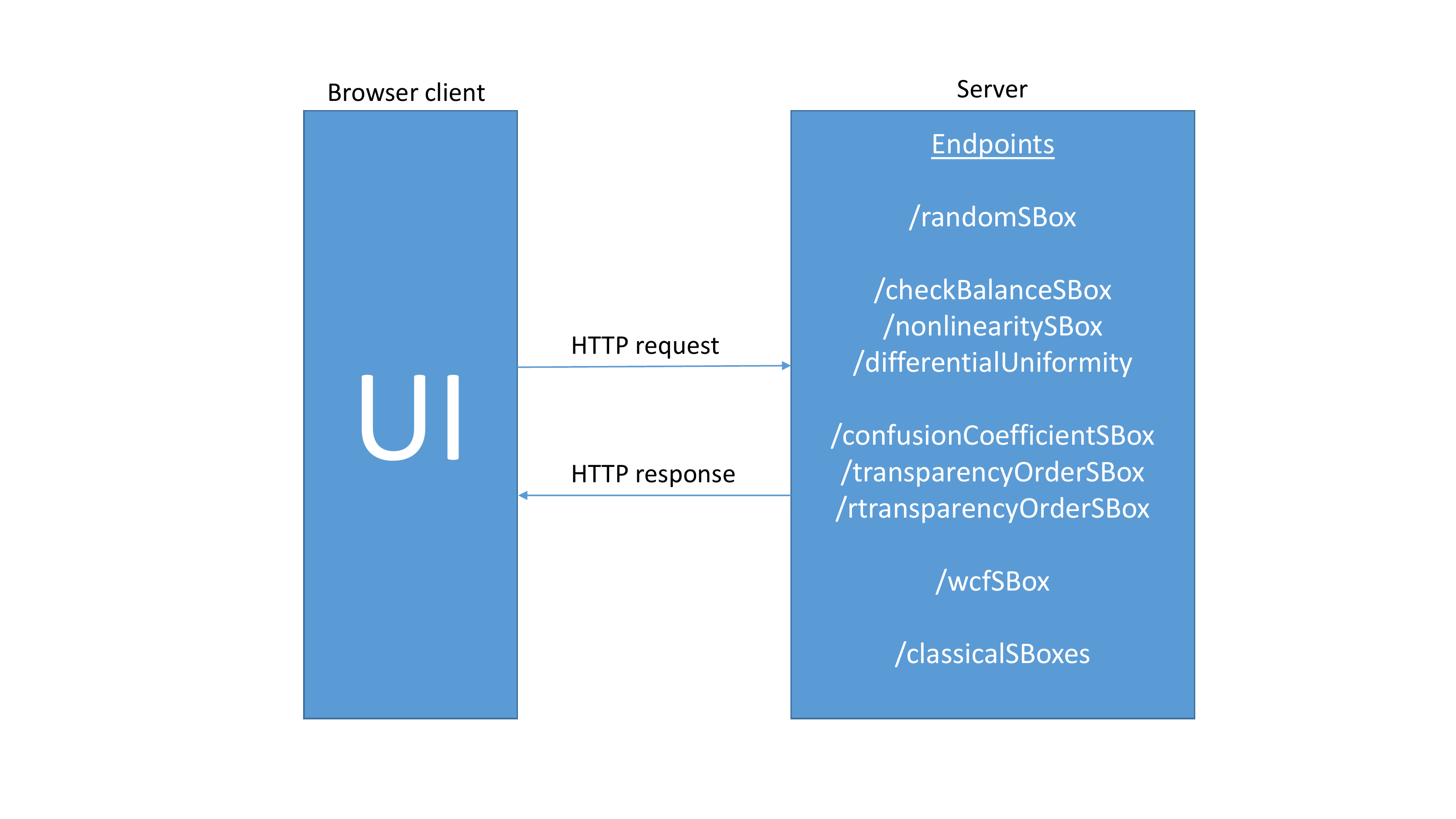}
	\caption{{\bf Main architecture.}
		User interface on browser client communicates to the server's endpoints by HTTP request/response scheme.}
	\label{fig1}
\end{figure}

The Ceibaco web application is hosted under the url \url{https://www.ceibaco.nat.cu/}. The main architecture of the application, Fig~\ref{fig1}, consists of a REST API backend (\textbf{PHP} code) and a Single Page Application frontend (\textbf{React/TypeScript} code). The backend contains one endpoint for 8x8 bijective S-boxes random generation, and others endpoints for calculate general bijective S-boxes properties. Also in the backend there is an endpoint to get a very small dataset of classical S-boxes.

The endpoints for properties' calculations should receive a JSON with the S-box to evaluate and the size of it (N and M). The response is also a JSON with the value of property evaluation.

\begin{figure}[!h]
	\includegraphics[width=\linewidth]{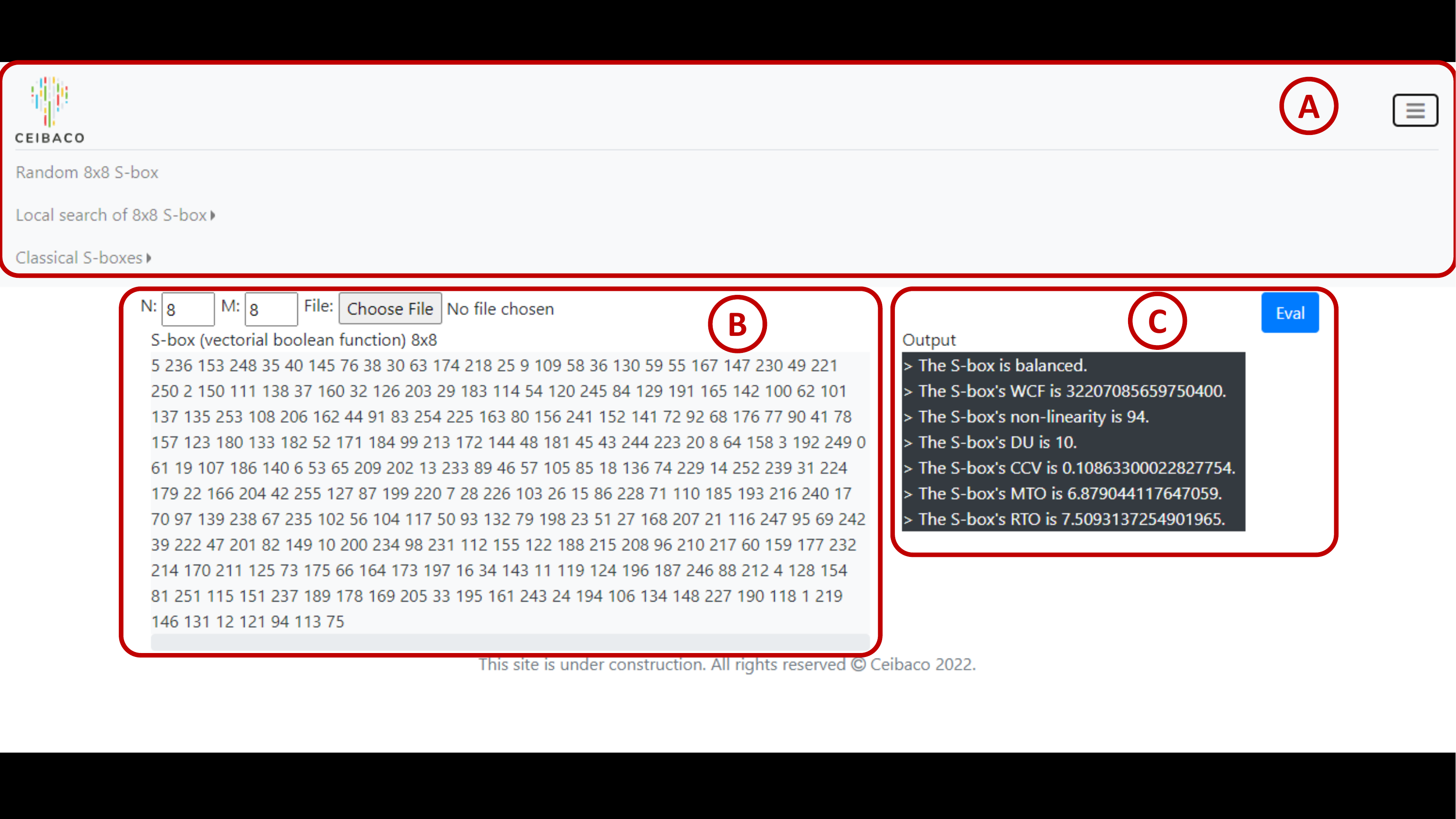}
	\caption{{\bf Main view.}
		Application contains three components. A: Menu component. B: S-box's loader component. C: S-box's evaluator component.}
	\label{fig2}
\end{figure}

In the Fig~\ref{fig2} we show the main view of the application. There is a menu for the application's options, a component for upload the S-box from file and another for evaluate the S-box's properties.

\subsection*{Dataset of classical S-boxes}

Our dataset Table~\ref{table1} contains four classical S-boxes that could be used to check the properties values. We select S-boxes of different size to test the correct properties calculations. These S-boxes can be recovered by REST API using the \textbf{/classicalSBoxes.php} endpoint.

\begin{table}[!ht]
	\centering
	\caption{
		{\bf Classical S-boxes.}}
	\begin{tabular}{|l|c|c|c|c|}
		\hline
		{\bf S-box} & {\bf NxN } & {\bf NL }& {\bf DU }& {\bf Ref. }\\ \thickhline
		AES & 8x8 & 112 & 4 & \cite{nyberg1992construction} \\ \hline
		KASUMI & 7x7 & 56 & 2 & \cite{kasumi} \\ \hline
		PRESENT & 4x4 & 4 & 4 & \cite{bogdanov2007present} \\ \hline
		PRINCE & 4x4 & 4 & 4 & \cite{borghoff2012low} \\ \hline
	\end{tabular}
	\begin{flushleft} These S-boxes have good values for Non-Linearity and Differential Uniformity.
	\end{flushleft}
	\label{table1}
\end{table}

\subsection*{API experiments}

Using the given endpoints, see Fig~\ref{fig1}, the developers can define experiments for search S-boxes with good values in its properties. The REST API can be accessible from any programming language through an HTTP library.

\begin{figure}[!h]
	\includegraphics[width=\linewidth]{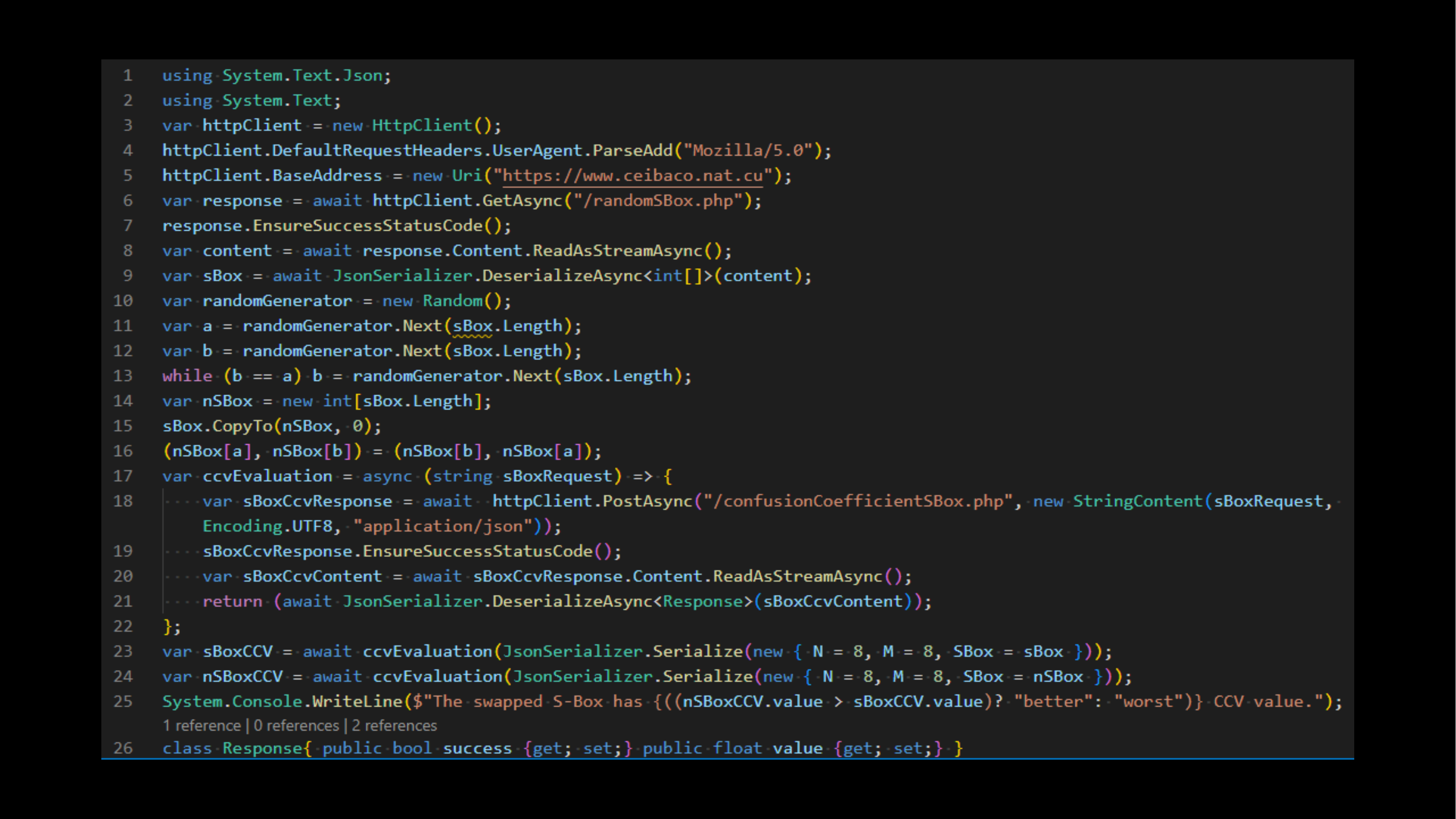}
	\caption{{\bf C\# code.}
		Experiment for compare the CCV value of a swapped S-box.}
	\label{fig3}
\end{figure}

As we mention before, the developer can define their own experiment using their preferred programming language. Fig~\ref{fig3} shows how to communicate to the REST endpoints. The \textbf{C\#} code in .NET 6 platform describe the random generation of an S-box, the creation a swapped S-box, the evaluation of CCV property for both S-boxes, and finally, the comparison of those values.

\subsubsection*{Testing}

\begin{figure}[!h]
	\includegraphics[width=\linewidth]{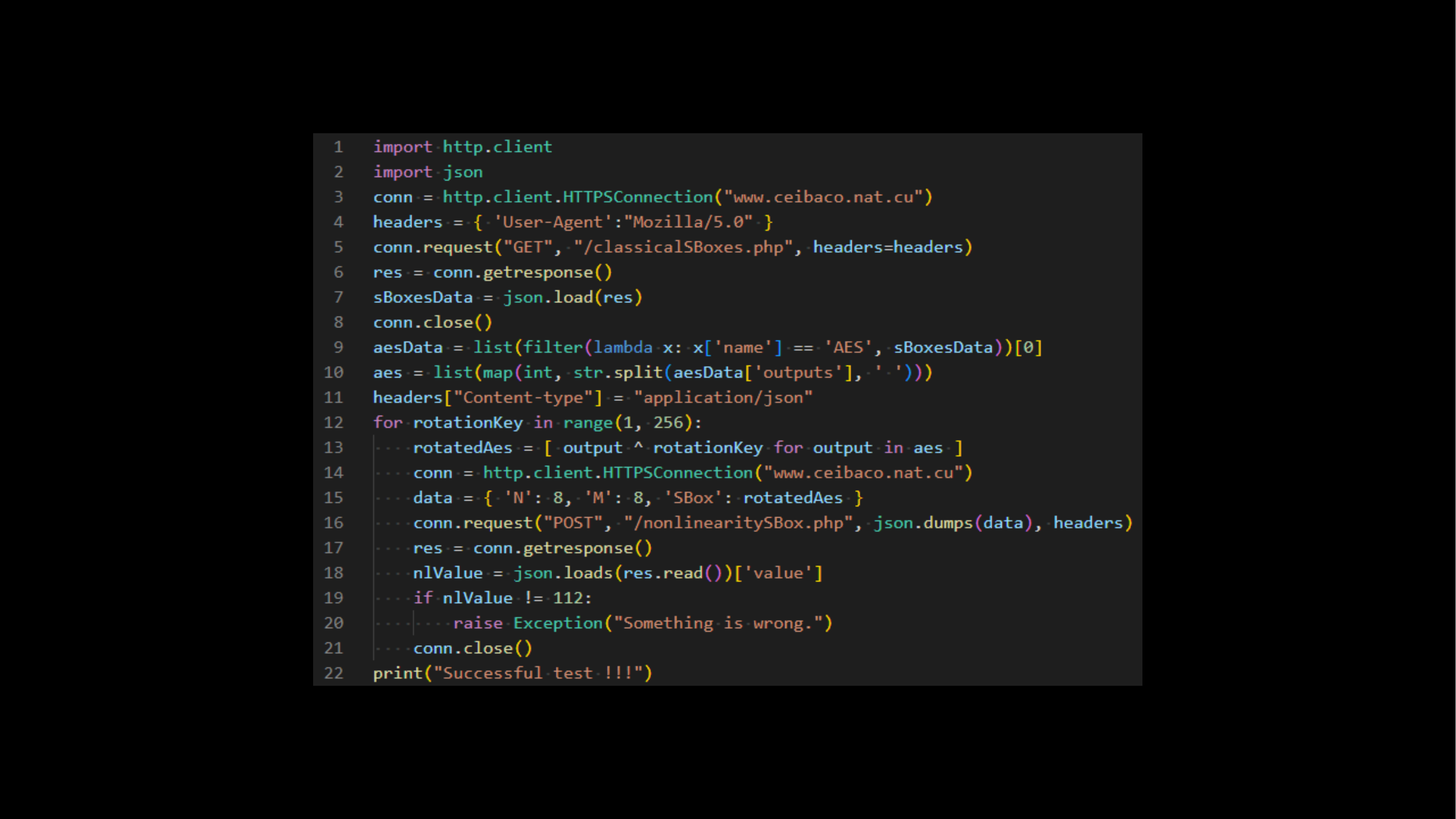}
	\caption{{\bf Python code.}
		Experiment for testing the NL value of AES's rotation S-boxes.}
	\label{fig4}
\end{figure}

The effectiveness of the application can be also tested using API experiments. In Fig~\ref{fig4} we define a \textbf{Python} code for evaluate the NL of S-boxes obtained by AES's rotation, those S-boxes should have a NL value of 112.

\subsection*{Local Search experiments}

Using the \textbf{TypeScript} language in our proposed SPA tool, we define two local search experiments, which can be executed by options in the menu component. The experiments follow the same local search algorithm presented in \cite{freyre2020external}, a fast algorithm for find S-boxes with high NL values that relies in the WalshHadamard Cost Function (WCF, \textbf{/wcfSBox.php} endpoint). The S-boxes obtained can have 100 or 102 of NL, depending of which option the user takes.

\begin{figure}[!h]
	\includegraphics[width=\linewidth]{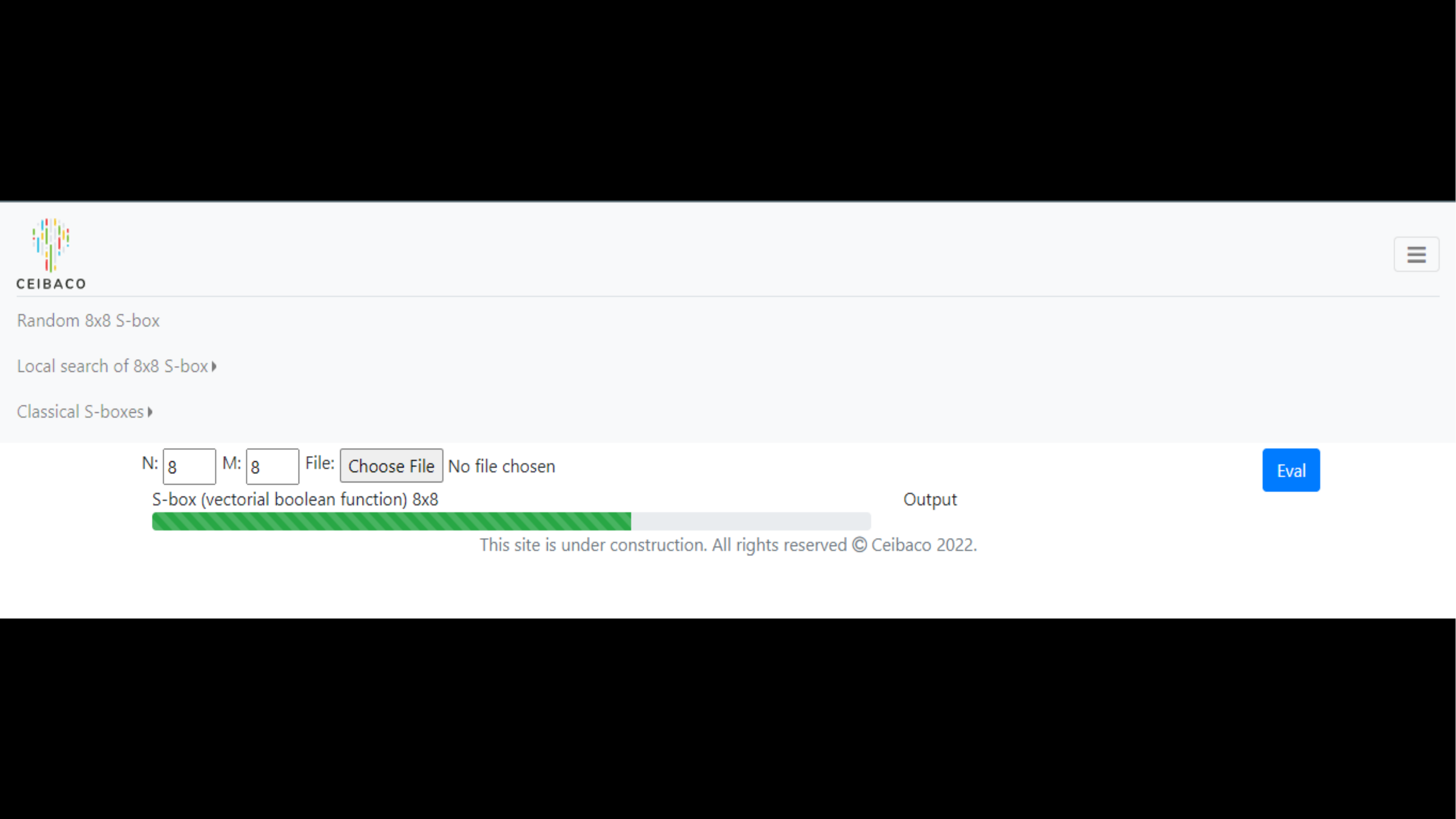}
	\caption{{\bf Progress of the local search experiment.}}
	\label{fig5}
\end{figure}

When the experiment is running, the user can check the progress of it in a bar, see Fig~\ref{fig5}. The progress represents how much close is the NL value of the actual solution with the expected NL value (100 or 102).

\section*{Discussion}

The web application architecture is different of the other tools implemented in \textbf{C/C++}. The properties evaluation is slower because the network calls and the use of \textbf{PHP} - an interpreted language -. We think some other languages should be consider for backend, like \textbf{C\#} of \textbf{Java}. Also, some properties and representations could be solved in the client. However, the benefits of using a REST API architecture is not negotiable; the API experiments allows developers do whatever they want as we show in the Results section. Local search experiments can be seen as an API experiment, but it is included inside the application.

In the case of application's graphical components, the three we define: menu, loader and evaluator, seems to be the essential components for this first software iteration. We really miss loading and evaluate more than one S-box, and this feature will be a top priority in the next version.

Although the classical dataset of S-boxes is too small, we decide to let it this way because there are not too much research about which S-boxes are very relevant; this important decision must take into account the relevance in literature and industry, good properties values and even the S-box size.

\section*{Conclusion}

The first REST API for the generation and evaluation of bijective S-boxes and the first SPA tool is presented. We also provide the way to define API experiments.

New directions to improve the tool could be including more properties to evaluate and other ways of generation. Processing multiples S-boxes is desirable.

\nolinenumbers

%
%
%
%
%
%
%

\bibliography{sample}

\end{document}